\documentclass[onecolumn, draftcls, 11pt]{IEEEtran}
\usepackage{amsmath, amssymb, amsfonts}
\usepackage{graphicx}
\usepackage{epsfig}
\usepackage{subfigure}
\usepackage[font={small}]{caption}
\usepackage{cases}
\usepackage{tensor}

\usepackage{color}
\usepackage{epstopdf}
\usepackage{balance}

\usepackage{tikz}
\usepackage{cite}
\usepackage{tensor}
\usepackage{mathtools}
\usepackage{empheq}

\makeatletter
\newcommand{\vast}{\bBigg@{4}}
\newcommand{\Vast}{\bBigg@{5}}
\makeatother

\begin{document}

\title{Stochastic Geometry-Based Performance Bounds for Non-Fading and Rayleigh Fading Ad Hoc Networks}
\author{ Hieu Duy Nguyen, \emph{Member, IEEE,} and Sumei Sun, \emph{Senior Member, IEEE}
\thanks{Manuscript received ...; revised ... ; accepted ... . This work was presented in part at ... . The associate editor coordinating the review of this paper and approving it for publication was ... .}
\thanks{The authors are with the Institute for Infocomm Research (I$^2$R), the Agency for Science, Technology and Research (A$\star$STAR), Singapore (email(s): \{nguyendh, sunsm\}@i2r.a-star.edu.sg).}
\thanks{Color versions of one or more of the figures in this paper are available online at http://ieeexplore.ieee.org.}
\thanks{Digital Object Identifier XXX.} \vspace{-0.1in}}

\maketitle


\newtheorem{definition}{\underline{Definition}}
\newtheorem{fact}{Fact}
\newtheorem{conjecture}{Conjecture}
\newtheorem{assumption}{\underline{Assumption}}
\newtheorem{theorem}{\underline{Theorem}}
\newtheorem{lemma}{\underline{Lemma}}
\newtheorem{corollary}{\underline{Corollary}}
\newtheorem{proposition}{\underline{Proposition}}
\newtheorem{example}{\underline{Example}}
\newtheorem{remark}{\underline{Remark}}
\newtheorem{algorithm}{\underline{Algorithm}}
\newcommand{\mv}[1]{\mbox{\boldmath{$ #1 $}}}

\captionsetup[table]{labelsep=newline}

\newcommand{\tikzcircle}[2][black,fill=black]{\tikz[baseline=-0.5ex]\draw[#1,radius=#2] (0,0) circle ;}%

\begin{abstract}
In this paper, we study the performance of non-fading and Rayleigh fading ad hoc networks. We first characterize the distribution of the signal-to-interference-plus-noise ratio (SINR) through the Laplace transform of the inverted SINR for non-fading channels. Since most communication systems are interference-limited, we also consider the case of negligible noise power, and derive the upper and lower bounds for the signal-to-interference ratio (SIR) distribution under both non-fading and fading cases. These bounds are of closed forms and thus more convenient for theoretical analysis. Based on these derivations, we obtain closed-form bounds for both the average Shannon and outage rates. We also leverage the above results to study partial fading ad-hoc systems. These results are useful for investigating and comparing fifth-generation communication systems, for example massive multi-antenna and small-cell networks as in our illustrative example.
\end{abstract}

\section{Introduction}\label{sec:intro}

Due to frequency reuse, a wireless communication system is affected by not only the thermal noise but also interference from other transmitters in the same frequency band. The thermal noise can be modeled as additive white Gaussian noise (AWGN). On the other hand, the network interference characteristic is often elusive as it depends on many aspects, for example the spatial structure of the overall network or the propagation characteristics. For ease of analysis, interference can be approximated by AWGN with some given power \cite{ViJa75}. However, such simple model does not capture many physical parameters affecting the interference. More realistic models, using hexagonal or square lattice, are often difficult to analyze and thus mostly rely on simulations for investigation. System modeling based on stochastic geometry can account for all practical parameters while yielding explicit performance expressions, hence is promising. 

Stochastic geometry has been applied to investigate communication networks as early as 1970s \cite{MuWa78}. It has been successfully integrated and adapted to cellular, ultra-wideband, cognitive radio, relay, and especially ad hoc networks \cite{StKeMe96,ChHa01,YaPe03,PiGiChWi09,GhSo08,ZhAn08,DoFrTh06}. Since wireless systems is mostly interference-limited, a majority of the works have focused on characterizing the interference and the signal-to-interference ratio (SIR) at each receiver/user equipment (UE). Please refer to \cite{BaBla09,HaAnBaDoFr09,Ha12} for detailed summaries. Under most networks, the interference and/or SIR distribution were represented only through Laplace transforms, which are often  complicated. Useful bounds of the SIR distribution for several system models have been proposed in \cite{WeAnJi07} and \cite{JiWeAn08}. In the seminal work \cite{AnBaGa11}, the fading ad hoc network has been investigated under the assumption that the desired signal power at a typical receiver/UE is exponentially distributed. Based on the Laplace transform and probability generating functionals of Poisson point processes (PPPs), Andrews \emph{et. al.} have shown that the interference and SIR distribution of a wide range of networks can be greatly simplified. Such approach has enabled simpler derivations of the outage probability and achievable rate, and thus motivated more investigations \cite{LeHu12,ZhHa14,JoNgSu14,SiAn14,GuDhViAn14,DeZhHa15}. 

Stochastic geometry is particularly useful in modeling and studying heterogeneous networks, which consists of many coexisting communication systems and in which all participating transmitters and receivers are randomly located \cite{SiAn14,GuDhViAn14,DeZhHa15}. Coordination and cooperation between cells have also been considered, e.g., in \cite{ZhHa14}. A shortcoming of many existing works is that they assume a very large UE density compared to the transmitter/BS density, effectively considering severely interfering networks only. To address this shortcoming, several recent studies have introduced a general model which includes the UE density as a varying parameter for performance investigation \cite{LeHu12,JoNgSu14}. 

In this paper, we obtain new results for the signal-to-interference-plus-noise ratio (SINR) and SIR cumulative distribution function (CDF) at a typical UE under non-fading and Rayleigh fading ad hoc networks, where the transmitting nodes are distributed according to PPPs. Particularly for non-fading ad hoc network, the Laplace transform of the inverted SINR are given, with which the distribution can be numerically calculated. We also derive upper- and lower bounds for the SIR CDF, which possess closed-form expressions and are thus more useful for subsequent analyses. For Rayleigh fading ad hoc network, we obtain closed-form bounds for the SIR CDF. Based on these results, the average Shannon and outage rate bounds are acquired under both cases. Similar results are obtained for partial fading ad-hoc systems. Since many communication networks can be modeled or reduced to these two ad hoc systems, our results are potentially useful for a wide range of problems. For example, we will show that our results can be applied to study and compare the performance of two important fifth-generation (5G) technologies, i.e., massive multiple-input multiple-output (M-MIMO) and small-cell densification. More information can be found in our subsequent paper \cite{NgSu15}.

The rest of the paper is organized as follows. Section \ref{sec:ad hoc netw} introduces the ad hoc network and its system model based on stochastic geometry. In Section \ref{sec:ResStoGeo NonFad}, we derive the Laplace transform of the inverted SINR, and the bounds of the SIR distribution and rates for non-fading ad hoc networks. Similar results for Rayleigh fading and partial fading ad hoc systems are obtained in Sections \ref{sec:ResStoGeo Fad} and \ref{sec:ResStoGeo ParFad}, respectively. Section \ref{sec:app and discuss} presents one application of our results to study M-MIMO performance and to compare it with that of small-cell densification. Finally, Section \ref{sec:conclusions} concludes the paper.

{\it Notations}: Scalars and vectors/matrices are denoted by lower-case and bold-face lower-case/upper-case letters, respectively. The conjugate, transpose, and conjugate transpose operators are denoted by $(\cdot)^*$, $(\cdot)^T$, and $(\cdot)^H$, respectively. $[\mv{A}]_{i,j}$ stands for the $(i,j)$th element of the matrix $\mv{A}$. $\mathbb{E}_{X}[\cdot]$ denotes the statistical expectation over a random variable $X$. $\mathrm{Tr}(\cdot)$, $||\cdot||_F$, and $\det (\cdot)$ represent the trace, Frobenius norm, and determinant of a matrix, respectively. $\mv{I}_{M}$ denotes an $M$-by-$M$ identity matrix. $\mathbb{R}^{x \times y}$ and $\mathbb{C}^{x \times y}$ denote the space of $x$-by-$y$ real and complex matrices, respectively. Finally, the circularly symmetric complex Gaussian distribution with mean $\mu$ and variance $\sigma^2$ is represented by $\mathcal{CN}(\mu,\sigma^2)$.

\section{System Model for Ad-Hoc Networks }\label{sec:ad hoc netw}
In this paper, we consider \emph{ad hoc} wireless networks with an infinite number of base stations (BSs) randomly distributed over an infinite area. The locations of the BSs are modeled by a homogeneous PPP $\Phi \subset \mathbb{R}^2$ with density $\lambda$. The single-antenna UEs are assumed to be distributed according to some independent stationary point process, and each UE is associated with the nearest transmit node. Furthermore, the UE density is large enough such that the BSs all transmit at any given time. The resulting system model is the so-called ad hoc network in the literature \cite{BaBla09,HaAnBaDoFr09,Ha12,WeAnJi07,JiWeAn08,AnBaGa11}. Consider an arbitrary BS $b$. Under the power-law path loss model, the base-band received signal for the UE in cell $b$ is given as
\begin{align}
y_b = h_{b,b} d_{b,b}^{-\mu/2} x_b + \sum_{k\neq b}^{\infty} h_{b,k} d_{b,k}^{-\mu/2} x_k + n_b,
\end{align}
where $d_{b,k}$ and $h_{b,k}$ are the distance and the independent and identically distributed (i.i.d.) fading coefficient between BS $k$ and the UE associated with BS $b$; $x_k$ is the transmit signal of BS $k$, where $\mathbb{E} \left[ |x_k|^2 \right] = P_T$ is the transmit power; $n_b$ is the additive white Gaussian noise (AWGN), $n_b \sim \mathcal{CN}(0,\sigma^2)$; and $\mu$ is the path loss exponent.

\begin{figure}[t]
 		\centering 
 		\epsfxsize=0.48\linewidth
 		\captionsetup{width=0.48\textwidth} 
 		\includegraphics[width=7.5cm, height=7.5cm]{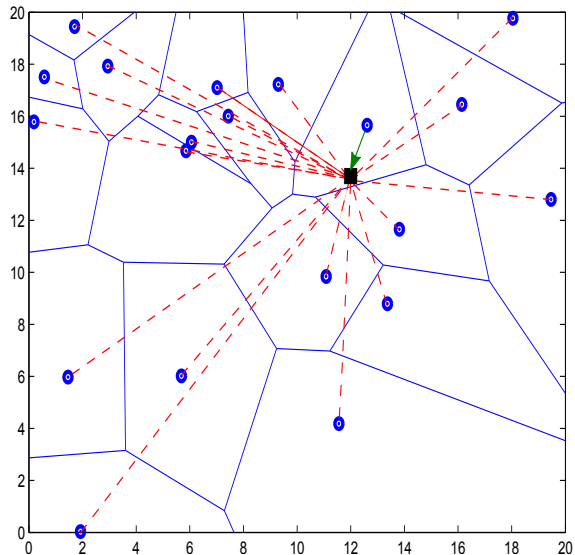}
 		\caption{A channel realization of an ad hoc network. Here, ``$\tikzcircle[fill=blue]{2.5pt}$'', ``$\blacksquare$'', and straight line denote the BSs, UE, and boundaries between the Voronoi cells. Furthermore, ``$\xrightarrow{}$'' and ``$- -$'' denote the channels between UE and its associated and interfering BSs, respectively. For ease of presentation, only one UE in one cell is shown here. }\label{fig:AdHoc ChRel}
\vspace{-0.1in}
\end{figure}

Since the system model is homogeneous, all analysis for a typical UE is applicable to any other user due to the Slivnyak's theorem \cite{BaBla09,HaAnBaDoFr09,Ha12}. For ease of notation, hereafter we consider the UE associated with BS $1$ and drop the index $1$ for brevity. The SINR for a typical UE can thus be given as
\begin{align}
\text{SINR} = \frac{\left| h_1 \right|^2 d_1^{-\mu}}{\sum_{k=2}^{\infty} \left| h_k \right|^2 d_k^{-\mu} + \delta},
\end{align}  
where $\delta \triangleq \sigma^2/P_T$ is the normalized noise power. Fig. \ref{fig:AdHoc ChRel} shows a channel realization of an ad hoc network with BS density $\lambda = 0.05$ km$^{-2}$ and service area assumed to be $20\times 20$ km$^2$.

\section{Performance of Non-Fading Ad-Hoc Networks}\label{sec:ResStoGeo NonFad}
In this section, we consider a non-fading ad hoc network where $\left| h_k \right|^2 = 1$, $\forall k$, in the system model described in Section \ref{sec:ad hoc netw}. Such system is also referred to as \emph{the base-line model} \cite{WeAnJi07}. We first obtain the Laplace transform of the inverted SINR. However, the resulting expression is complicated and thus difficult to use for theoretical analysis. We then derive the upper and lower bounds for the SIR distribution, which are in closed forms. Based on these results, we investigate and obtain the bounds for both the average Shannon and the outage rates of non-fading ad hoc networks.

Consider the random variable 
\begin{align}\label{eq:SINR Q}
Q_{\text{n-fd}} \triangleq \frac{d_1^{-\mu}}{\sum_{k=2}^{\infty} d_k^{-\mu} + \delta}.
\end{align}

It represents the SINR at a typical UE under non-fading ad hoc networks with noise power $\delta$. However, existing works have assumed that the distance $d_{1}$ is a constant and only the set $\left\{ d_{k} \right\}_{k=2}^{\infty}$ follows a PPP \cite{WeAnJi07,WeAnJi10}. In contrast, we consider the case when $\left\{ d_{k} \right\}_{k=1}^{\infty}$ follows the PPP $\Phi$ with density $\lambda$ and $d_k \geq d_1$, $\forall k \geq 2$. This is a typical regime where the UE is associated with the nearest transmitter. The distance from the UE to the associated transmitter, $d_1$, is therefore smaller than those to other transmitters, $\left\{ d_k \right\}_{k=2}^{\infty}$.

We first derive the Laplace transform of $Q_{\text{n-fd}}$, based on which the distribution of $Q_{\text{n-fd}}$ can be calculated numerically. We furthermore propose CDF upper and lower bounds of $Q_{\text{n-fd}}$, which is much more useful for theoretical analysis than the Laplace transform. 

\begin{theorem}\label{theorem:GenLap}
The Laplace transform of the random variable $Q_{\text{n-fd}}^{-1}$ is given as
\begin{align}\label{eq:GenLap}
\mathcal{L}_{1/Q_{\text{n-fd}}} \left(s \right) 
& = \pi \lambda \int_{0}^{\infty}  \exp \bigg\{ -s y^{\mu/2} \delta - \pi \lambda y \notag \\
& \qquad \times \left[ 2- e^{-s} + s^{2/\mu} \gamma\left( 1- \frac{2}{\mu},s \right)  \right]   \bigg\} dy,
\end{align}
\end{theorem}
\begin{IEEEproof}
Please refer to Appendix \ref{appen:GenLap}.
\end{IEEEproof}

Theorem \ref{theorem:GenLap} provides a semi-analytical approach to calculate the distribution of $Q_{\text{n-fd}}^{-1}$, and consequently of $Q_{\text{n-fd}}$. However, such a characterization requires intensive numerical computations, but does not provide any insightful observation. We thus seek to find bounds for the CDF of $Q_{\text{n-fd}}$ which is more useful for theoretical analyses. Here we assume that the noise power $\delta$ is small and approximate $Q_{\text{n-fd}} \approx \frac{d_1^{-\mu}} {\sum_{k=2}^{\infty} d_k^{-\mu}}$. In M-MIMO systems, this approximation is well justified due to effect of M-MIMO beamforming which compels the noise power to much smaller than the inter-BS interference. The following lemma establishes upper and lower bounds of the CDF of $Q_{\text{n-fd}}$ given that $\delta = 0$.
\begin{lemma}\label{lem:BouCDF g1}
Given that $\delta = 0$ and $q\geq 1$, the CDF $F_{Q_{\text{n-fd}}}(q) \triangleq \text{Pr} \{ Q_{\text{n-fd}} \leq q \}$ is upper- and lower-bounded as $F_{Q_{\text{n-fd}}}^{\text{LB}}(q)  \leq$ $F_{Q_{\text{n-fd}}}(q) \leq$ $F_{Q_{\text{n-fd}}}^{\text{UB}}(q)$, where
\begin{align}
F_{Q_{\text{n-fd}}}^{\text{LB}}(q) 
& = 1 - \frac{1}{q^{2/\mu}}, \label{eq:LB Q g1} \\
F_{Q_{\text{n-fd}}}^{\text{UB}}(q) 
& = 1 -  \frac{1}{\beta_{\text{n-fd}} q^{2/\mu}}, \label{eq:UB Q g1}
\end{align}
and 
\begin{align}
\beta_{\text{n-fd}} = \frac{(\mu+2)^{2/\mu+1}}{\mu (\mu-2)^{2/\mu}}.
\end{align}
\end{lemma}
\begin{IEEEproof}
Please refer to Appendix \ref{appen:BouCDF g1}.
\end{IEEEproof}

It is difficult to investigate the regime $q < 1$. In the following, we provide an asymptotic upper bound for $Q_{\text{n-fd}}$ as $q \to 0$ or $q \to 1$.
\begin{lemma}\label{lem:BouCDF l1}
Given that $\delta = 0$ and $q < 1$ but $q \to 0$ or $1$, the CDF $F_{Q_{\text{n-fd}}}(q)$ is upper-bounded as
\begin{align}\label{eq:UB Q l1}
F_{Q_{\text{n-fd}}}^{\text{UB}}(q) = 1 -  \frac{1}{1 + \beta_{\text{n-fd}} q^{2/\mu}},
\end{align}
\end{lemma}
\begin{IEEEproof}
Please refer to Appendix \ref{appen:BouCDF l1}.
\end{IEEEproof}

Based on Lemmas \ref{lem:BouCDF g1} and \ref{lem:BouCDF l1}, we propose the following universal bounds for $Q_{\text{n-fd}}$.
\begin{subnumcases}{\label{eq:LB all Q} F_{Q_{\text{n-fd}}}^{\text{LB}}(q) = }
0, &  $0\leq q \leq 1$, \label{case:LBqLeq1} \\
1 - \frac{1}{q^{2/\mu}}, &  $q > 1$, \label{case:LBqGeq1}
\end{subnumcases} 
\begin{align}\label{eq:UB all Q}
F_{Q_{\text{n-fd}}}^{\text{UB}}(q) = 1 -  \frac{1}{1 + \beta_{\text{n-fd}} q^{2/\mu}}. ~~ \quad \qquad \quad 
\end{align}

\begin{figure}[t]
 		\centering 
 		\epsfxsize=0.48\linewidth
 		\captionsetup{width=0.48\textwidth} 
 		\includegraphics[width=8.5cm, height=7.5cm]{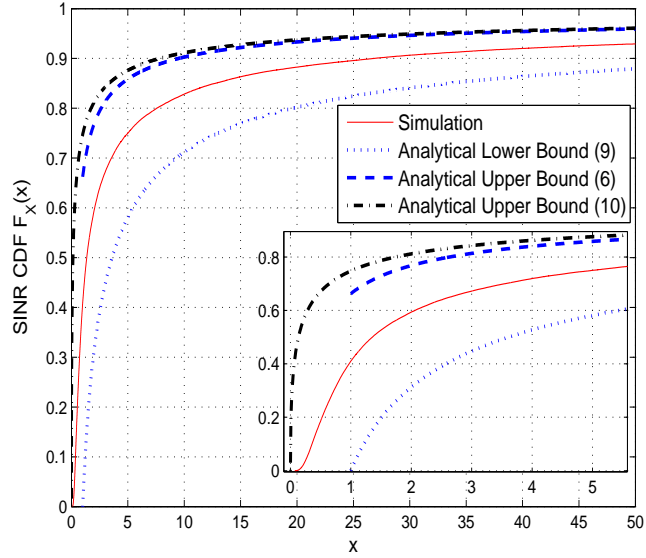}
 		\caption{Simulated SINR CDF and its bounds for non-fading ad hoc network with $\mu=3.7$ and $\delta=-20$ dB.}\label{fig:CDF non-fad}
\vspace{-0.1in}
\end{figure}

\begin{figure}[t]
 		\centering 
 		\epsfxsize=0.48\linewidth
 		\captionsetup{width=0.48\textwidth} 
 		\includegraphics[width=8.5cm, height=7.5cm]{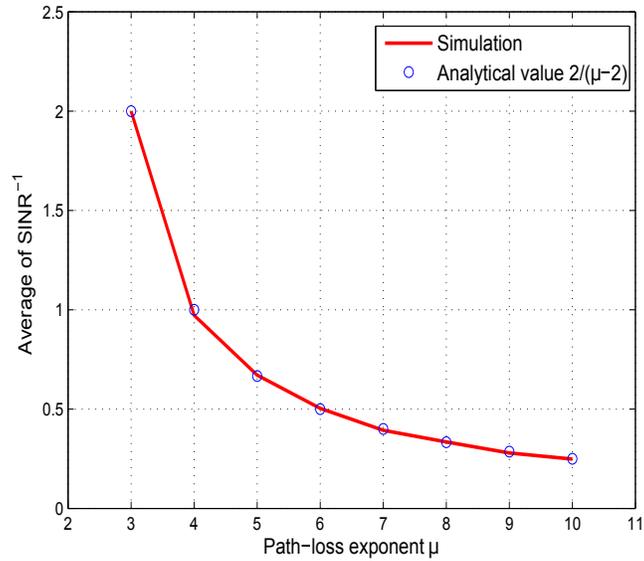}
 		\caption{Average of $Q_{\text{n-fd}}^{-1}$ with respect to the path loss exponent.}\label{fig:Avg SINR Inv non-fad}
\vspace{-0.1in}
\end{figure}

Analytically, (\ref{eq:UB all Q}) has been proved only when: $q\geq 1$; $q\to 0$; or $q<1$ but $q\to 1$ in Lemmas \ref{lem:BouCDF g1} and \ref{lem:BouCDF l1}. However, intensive simulations show that it holds for all values of $q$. In Fig. \ref{fig:CDF non-fad}, we show the simulated SINR CDF and the bounds given in (\ref{eq:UB Q g1}), (\ref{eq:LB all Q}), and (\ref{eq:UB all Q}), for an ad hoc network with $\mu=3.7$, $\delta=-20$ dB, and $20000$ channel realizations. We see that the bounds are close to the exact distribution. With a path-loss exponent following that of LTE, i.e., $\mu \in [3.7, 4]$ \cite{LTE01}, the difference between (\ref{eq:UB Q g1}) and the looser (but universal) (\ref{eq:UB all Q}) is negligible. However, as $\mu$ increases, we observe that (\ref{eq:UB all Q}) deviates farther from the exact CDF than (\ref{eq:UB Q g1}). Another observation is that the tail of the exact SINR CDF seems to follow a mixed scaling law $\sim x^{-a} \exp(-b x)$. This observation might help subsequent works for developing tighter CDF bounds.

The following Corollary \ref{corol:E Q^-1} gives the expectation of $Q_{\text{n-fd}}^{-1}$, averaged over the PPP $\Phi$. This result might be useful for comparison between different systems. The proof can be deduced from Appendix \ref{appen:BouCDF l1}. A confirmation for Corollary \ref{corol:E Q^-1} is illustrated in Fig. \ref{fig:Avg SINR Inv non-fad}, where we plot the inverted SINR versus the path-loss exponent with $\delta=-20$ dB $\approx 0$.
\begin{corollary}\label{corol:E Q^-1}
For non-fading channel, the expectation of the inverted SIR is given as $\mathbb{E}_{\Phi} \left[ Q_{\text{n-fd}}^{-1} \big| \delta = 0 \right] = \frac{2}{\mu-2}$. 
\end{corollary}

\subsection{Average Achievable Rate}\label{subsec:NonFad avg rate}

In this subsection, we obtain several results for the average Shannon rate of a typical user, defined as
\begin{align}
R_{\text{n-fd}}(\alpha) \triangleq \mathbb{E}_{\Phi} \left[ \log_2 \left( 1 + \alpha Q_{\text{n-fd}} \right) \right],
\end{align}
where $Q_{\text{n-fd}}$ is given in (\ref{eq:SINR Q}) and $\alpha$ is a constant representing, e.g., the beamforming gain of the system. Essentially, we have assumed that single-user receiver is employed, i.e., the interference is treated as noise. We first note the following fact 
\begin{fact}\label{fact:E to Pr}
For a positive random variable $X$, the expectation of $X$ can be expressed as $\mathbb{E} \left[ X \right] = \int_{0}^{\infty} \text{Pr} \left\{ X > t \right\} dt$.
\end{fact}

Based on Fact \ref{fact:E to Pr}, the average rate can be written as
\begin{align}\label{eq:R avg}
R_{\text{n-fd}}(\alpha) 
& = \int_{0}^{\infty} \text{Pr} \left\{ \log_2 \left( 1 + \alpha Q_{\text{n-fd}} \right) > t \right\} dt \notag \\
& = \int_{0}^{\infty} \text{Pr} \left\{ Q_{\text{n-fd}} > \frac{2^t - 1}{\alpha} \right\} dt.
\end{align}

Therefore, an exact $R_{\text{n-fd}}(\alpha)$ can be calculated numerically by using the Laplace transform of $Q_{\text{n-fd}}^{-1}$ in Theorem \ref{theorem:GenLap}. Such approach, however, does not give us any insight on the system performance. We therefore provide bounds of $R_{\text{n-fd}}(\alpha)$ based on (\ref{eq:LB all Q}) and (\ref{eq:UB all Q}) in the next lemma.
\begin{lemma}\label{lemma:Ravg UB LB}
The average Shannon rate $R_{\text{n-fd}}(\alpha)$ is upper- and lower-bounded by $R_{\text{n-fd}}^{\text{UB}}(\alpha)$ and $R_{\text{n-fd}}^{\text{LB}}(\alpha)$, respectively, where
\begin{align}
R_{\text{n-fd}}^{\text{UB}}(\alpha) 
& = \log_2(1+\alpha) + \frac{\alpha^{2/\mu} \mu}{(\mu-2)\log 2} \rho \left( \alpha^{1-2/\mu} , \frac{\mu}{\mu-2} \right), \label{eq:R UB non-fad} \\
R_{\text{n-fd}}^{\text{LB}}(\alpha) 
& = \int_{0}^{\infty} \frac{\alpha^{2/\mu} dt}{\alpha^{2/\mu} + \beta_{\text{n-fd}} \left( 2^t - 1 \right)^{2/\mu} }, \label{eq:R LB non-fad}
\end{align}
\end{lemma}
where $\rho(a,b)  \triangleq \int_{a}^{\infty} \frac{dx}{x^b+1}$. 
\begin{IEEEproof}
Please refer to Appendix \ref{appen:Ravg UB LB}.
\end{IEEEproof}

\subsection{Outage Rate}\label{subsec:NonFad outage rate}

\begin{figure}[t]
 		\centering 
 		\epsfxsize=0.48\linewidth
 		\captionsetup{width=0.48\textwidth} 
 		\includegraphics[width=8.5cm, height=7.5cm]{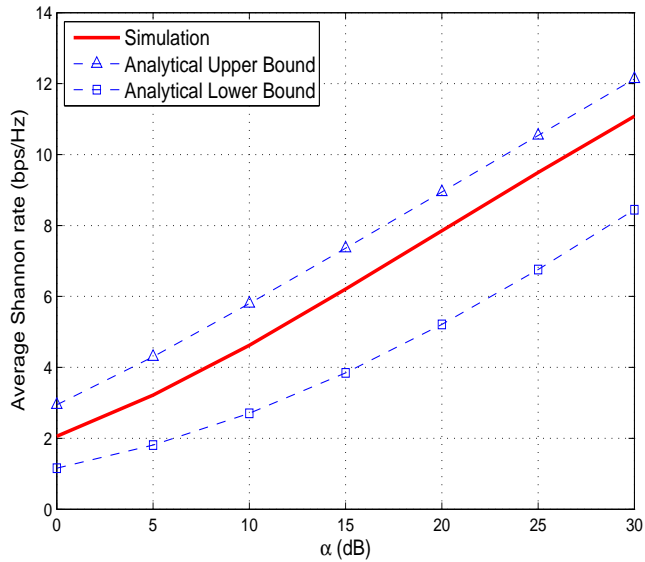}
 		\caption{Average Shannon rate and its bounds (\ref{eq:OR UB non-fad}), (\ref{eq:OR LB non-fad}) for non-fading ad hoc network with $\mu=3.7$.}\label{fig:Avg rate non-fad}
\vspace{-0.1in}
\end{figure}

Outage rate is another performance metric which for our system can be defined as
\begin{align}\label{eq:OR def}
\text{OR}_{\text{n-fd}}(\alpha, \eta) \triangleq \text{Pr} \left\{ Q_{\text{n-fd}} \geq \frac{\eta}{\alpha} \right\} \log_2 \left( 1 + \eta \right),
\end{align}
where $\alpha$ is a given beamforming gain and $\eta$ is the target SINR corresponding to a target rate of $\log_2(1+\eta)$. Similar to the Shannon throughput, the exact outage rate can be numerically determined based on the Laplace transform (\ref{eq:GenLap}). Again, we derive the bounds for the outage rate in the following lemma.
\begin{lemma}\label{lemma:OR UB LB}
The outage rate $\text{OR}_{\text{n-fd}}(\alpha,\eta)$ with $\alpha \geq 0$ is upper- and lower-bounded by $\text{OR}_{\text{n-fd}}^{\text{UB}}(\alpha,\eta)$ and $\text{OR}_{\text{n-fd}}^{\text{LB}}(\alpha,\eta)$, respectively, where
\begin{subnumcases}{\label{eq:OR UB non-fad} \text{OR}_{\text{n-fd}}^{\text{UB}}(\alpha,\eta) = }
\log_2(1+\eta), & \hspace{-0.3in} $0 \leq \eta \leq \alpha$, \label{case:OR UB etaSmall} \\
\log_2(1+\eta) \left( \frac{\alpha}{\eta} \right)^{2/\mu}, & \hspace{-0.3in} $\eta > \alpha$. \label{case:OR UB etaLarge}
\end{subnumcases}
\begin{flalign}\label{eq:OR LB non-fad}
\text{OR}^{\text{LB}}_{\text{n-fd}}(\alpha,\eta) & = \frac{\log_2 \left(1+\eta \right)}{1 + \beta_{\text{n-fd}} \left(  \frac{\eta}{\alpha} \right)^{2/\mu} }.   & 
\end{flalign}
where $\beta_{\text{n-fd}}$ is defined in Lemma \ref{lem:BouCDF g1}.
\end{lemma}
\begin{IEEEproof}
The proof is obtained by substituting (\ref{eq:LB Q g1}) and (\ref{eq:UB all Q}) into (\ref{eq:OR def}). It is thus omitted for brevity.
\end{IEEEproof}

The outage capacity of a system can be defined as the maximum of $\text{OR}_{\text{n-fd}}(\alpha,\eta)$ over $\eta$, i.e., $\text{OC}_{\text{n-fd}}(\alpha) = \max_{\eta \geq 0} \text{OR}_{\text{n-fd}}(\alpha,\eta)$. However, it is not easy to derive $\text{OC}_{\text{n-fd}}(\alpha)$ based on Lemma \ref{lemma:OR UB LB}.

\begin{figure}[t]
 		\centering 
 		\epsfxsize=0.48\linewidth
 		\captionsetup{width=0.48\textwidth} 
 		\includegraphics[width=8.5cm, height=7.5cm]{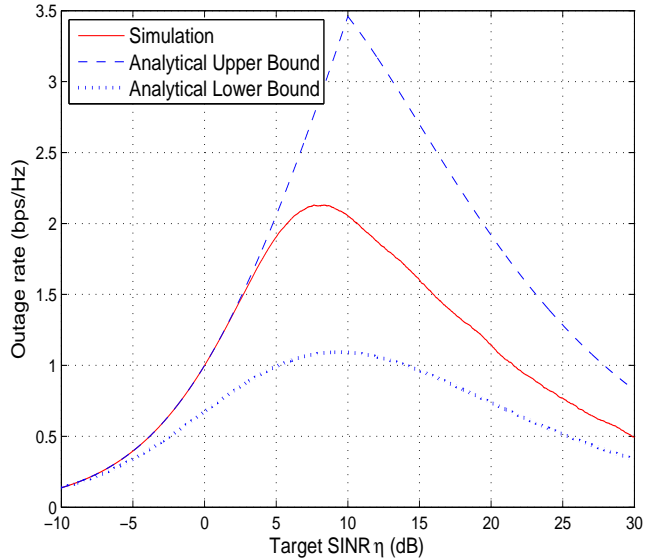}
 		\caption{Outage rate and its bounds (\ref{eq:R UB non-fad}), (\ref{eq:R LB non-fad}) for non-fading ad hoc network with respect to the target SINR. Here $\mu=3.7$ and $\alpha=10$ dB.}\label{fig:Outage rate non-fad}
\vspace{-0.1in}
\end{figure}

In Figs. \ref{fig:Avg rate non-fad} and \ref{fig:Outage rate non-fad}, we illustrate the average Shannon rate and the outage rate over $\alpha$ and $\eta$, respectively. For comparison, we also plot their analytical upper and lower bounds given in Lemmas \ref{lemma:Ravg UB LB} and \ref{lemma:OR UB LB}. It is seen that the upper bound is tighter than the lower bound for the average rate. The observation is reversed for the outage rate, in which the lower bound is (slightly) tighter. One reason for the slackness of the lower bound is due to the overestimation of (\ref{eq:UB all Q}). Intensive simulations suggest that the tail of the exact SINR CDF follows $\sim x^{-a} \exp(-b x)$, which approaches $0$ faster than the polynomial law $x^{-2/\mu}$ given by the current CDF bounds (\ref{eq:LB all Q}) and (\ref{eq:UB all Q}). To obtain tighter CDF bounds, more stringent analyses are necessary.

\begin{remark}\label{remark:1}
Assume that the noise power is ignored, i.e., $\delta=0$. From Theorem \ref{theorem:GenLap}, we observe that the Laplace transform of $Q_{\text{n-fd}}^{-1}$ does not depend on the density $\lambda$, and it can be expressed as
\begin{align}
\mathcal{L}_{1/Q_{\text{n-fd}}} \left(s \right) 
& = \int_{0}^{\infty}  \exp \bigg\{ - t  \bigg[ 2- e^{-s} \notag \\
& \qquad \qquad + s^{2/\mu} \gamma\left( 1- \frac{2}{\mu},s \right)  \bigg]   \bigg\} dt. \label{eq:GenLap delta=0}
\end{align}
\end{remark}

As a result, the CDF and rate of non-fading ad hoc networks are independent of the density. Such observation is readily demonstrated in Lemmas \ref{lem:BouCDF g1} to \ref{lemma:OR UB LB}. At first glance, it seems to contradict with the results in \cite{WeAnJi07,WeAnJi10}. The reason for the difference is that Weber \emph{et. al.} only assumed the interfering BS locations to distribute according to a PPP, but not the associated BS. The distance between the UE and associated BS is a constant in \cite{WeAnJi07,WeAnJi10}. However in this study, we assume that all BS locations are distributed according to a PPP.

We also note that the independence of the CDF and rate to the density can be explained geometrically. Having a higher density resembles the effect of zooming in a certain region and adding more transmitting nodes. Since the considered PPP is homogeneous, the resulting PPP is stochastically equivalent to the original one. The relative ratios between the distances are therefore unchanged. It explains why the CDF and the rate are unaffected by the BS density.  

\section{Performance of Fading Ad-Hoc Networks}\label{sec:ResStoGeo Fad}

In this section, we consider a slightly different problem than Section \ref{sec:ResStoGeo NonFad}. Particularly, we investigate the quantity
\begin{align}\label{eq:SINR Q fad}
Q_{\text{fd}} \triangleq \frac{\left| h_1 \right|^2 d_1^{-\mu}}{\sum_{k\in \Phi \setminus \{1\}} \left| h_k \right|^2 d_k^{-\mu} + \delta},
\end{align}
where $\Phi$, $\left\{ h_k \right\}_{k=1}^{\infty}$, $\left\{ d_k \right\}_{k=1}^{\infty}$, and $\delta$ have been defined in Section \ref{sec:ad hoc netw}. In the literature, $Q_{\text{fd}}$ is the SINR of an ad hoc network under fading channels, and also attracted much attention. The exact outage probability of $Q_{\text{fd}}$ is given in \cite{AnBaGa11} for the general and various special cases. The crucial assumption in \cite{AnBaGa11} is the exponential distribution of $\left| h_1 \right|^2$, which allows a simple representation of the SINR distribution. 

\begin{figure}[t]
 		\centering 
 		\epsfxsize=0.48\linewidth
 		\captionsetup{width=0.48\textwidth} 
 		\includegraphics[width=8.5cm, height=7.5cm]{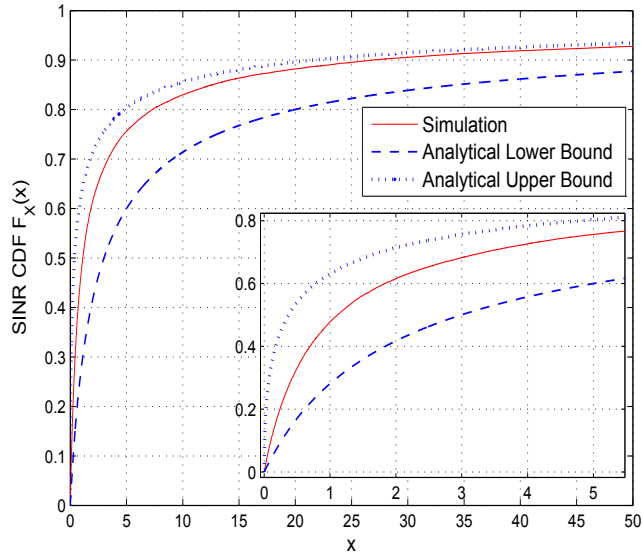}
 		\caption{Simulated SINR CDF and its bounds (\ref{eq:FD LB CDF exact}), (\ref{eq:FD UB CDF exact}) for fading ad hoc network with $\mu=3.7$ and $\delta=-15$ dB.}\label{fig:CDF fad}
\vspace{-0.1in}
\end{figure}

We are interested in the Rayleigh fading channel, i.e., $\left| h_k \right|^2$ is exponentially distributed with variance $1$, of which the SINR CDF is provided in \cite[Theorem 2]{AnBaGa11}. Assuming that the noise power is $\delta = 0$, a much simpler expression can be obtained as \cite{AnBaGa11}
\begin{align}\label{eq:SINR Q fad NoNoise}
F_{Q_{\text{fd}}}(q) = \text{Pr} \left\{ Q_{\text{fd}} \leq q \right\} = 1 - \frac{1}{1 + q^{2/\mu} \rho\left( q^{-2/\mu}, \frac{\mu}{2} \right)  },
\end{align}  
where the function $\rho(a,b)$ is defined in Lemma \ref{lemma:Ravg UB LB}. The expression (\ref{eq:SINR Q fad NoNoise}) requires a numerical integration and is thus not useful for performance analysis. We will develop the lower and upper bounds for (\ref{eq:SINR Q fad NoNoise}).
\begin{lemma}\label{lemma:FD UB LB CDF}
Assume $\delta = 0$. The lower and upper bounds for the CDF of $Q_{\text{fd}}$ defined in (\ref{eq:SINR Q fad}) are given as
\begin{align}
F^{\text{LB}}_{Q_{\text{fd}}}(q) 
& = 1 - \mathbb{E}_{h} \left[ \frac{1}{1 + \frac{2 q^{2/\mu}}{\mu |h|^{2/\mu} } \Gamma \left( \frac{2}{\mu}, \frac{|h|^2}{q} \right) } \right], \label{eq:FD LB CDF exact} \\
F^{\text{UB}}_{Q_{\text{fd}}}(q) 
& = 1 - \frac{1}{1 + \beta_{\text{fd}} q^{2/\mu} }. \label{eq:FD UB CDF exact}
\end{align}
where $h \sim \mathcal{CN}(0,1)$; and 
\begin{align}
\beta_{\text{fd}} \triangleq  \frac{2\pi}{\mu \sin \left( \frac{2\pi}{\mu} \right)} = \Gamma \left( 1 + \frac{2}{\mu} \right) \Gamma \left( 1 - \frac{2}{\mu} \right).
\end{align} 
\end{lemma}
\begin{IEEEproof}
Please refer to Appendix \ref{appen:FD UB LB CDF}.
\end{IEEEproof}

Fig. \ref{fig:CDF fad} depicts the SINR CDF and the bounds given in Lemma \ref{lemma:FD UB LB CDF} for $\mu=3.7$ and $\delta=-15$ dB. Unlike the non-fading network (see Fig. \ref{fig:CDF non-fad}), the bounds for the fading case is much stricter, especially the upper bound. For path-loss exponents under typical LTE regimes, i.e., $\mu=3.7$ to $4$ \cite{LTE01}, the upper bound is very close to the exact CDF, and gradually becomes looser as $\mu$ increases. It is expected that the gaps between the exact rates and bounds given in Lemma \ref{lemma:Rate FD} will be smaller than those under the non-fading case. 

In the following, we derive the average achievable Shannon and outage rate bounds based on the CDF upper bound. The proofs are similar to those in Sections \ref{subsec:NonFad avg rate} and \ref{subsec:NonFad outage rate}, and thus omitted for brevity. 
\begin{lemma}\label{lemma:Rate FD}
The average Shannon rate $R_{\text{fd}}(\alpha)$ and the outage rate $\text{OR}_{\text{fd}}(\alpha,\eta)$ are bounded above, respectively, by 
\begin{align}
R_{\text{fd}}^{\text{LB}}(\alpha) 
& = \int_{0}^{\infty} \frac{\alpha^{2/\mu} dt}{\alpha^{2/\mu} + \beta_{\text{fd}} \left( 2^t - 1 \right)^{2/\mu} }, \label{eq:R LB fad} \\
\text{OR}^{\text{LB}}_{\text{fd}}(\alpha,\eta)
& = \frac{\log_2 \left(1+\eta \right)}{1 + \beta_{\text{fd}} \left(  \frac{\eta}{\alpha} \right)^{2/\mu} }. \label{eq:OR LB fad}
\end{align}
\end{lemma}

\begin{figure}[t]
 		\centering 
 		\epsfxsize=0.48\linewidth
 		\captionsetup{width=0.48\textwidth} 
 		\includegraphics[width=8.5cm, height=7.5cm]{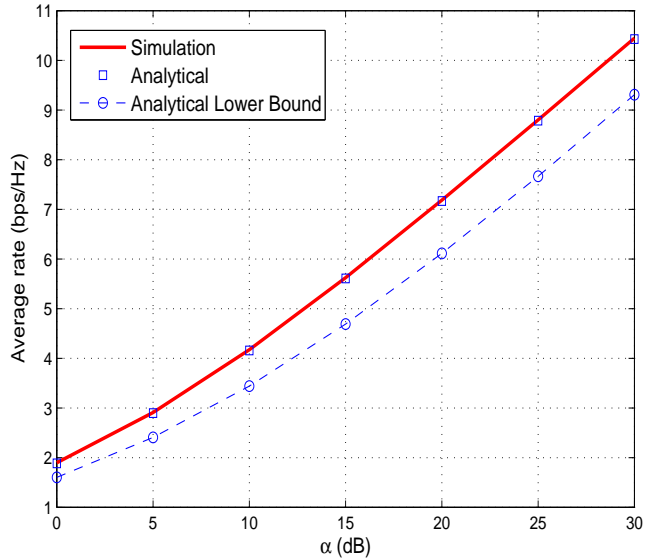}
 		\caption{Average Shannon rate and its bound (\ref{eq:R LB fad}) for Rayleigh fading ad hoc network with $\mu=3.7$ and $\delta = -15$ dB. The analytical values are calculated based on the noise-free approximation (\ref{eq:SINR Q fad NoNoise}). }\label{fig:Avg rate fad}
\vspace{-0.1in}
\end{figure}

The rate upper bounds can also be derived, however, they possess complicated integral forms with little practical use. The reason is due to (\ref{eq:FD LB CDF exact}) which requires numerical calculations. We therefore do not present the rate lower bounds here.

\begin{figure}[t]
 		\centering 
 		\epsfxsize=0.48\linewidth
 		\captionsetup{width=0.48\textwidth} 
 		\includegraphics[width=8.5cm, height=7.5cm]{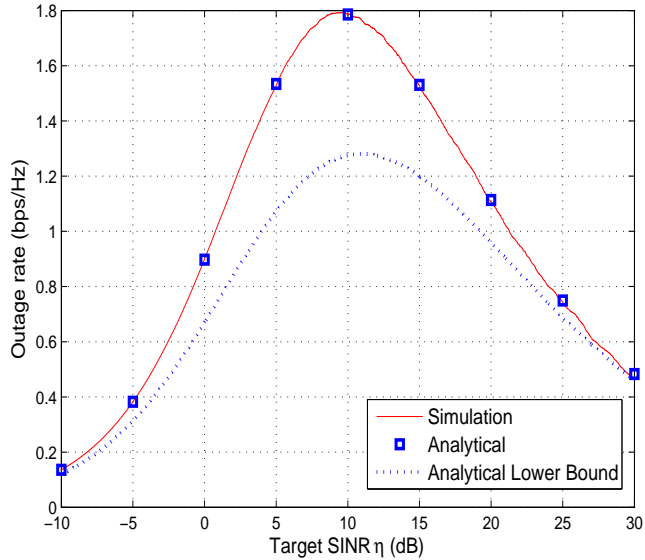}
 		\caption{Outage rate and its bound (\ref{eq:OR LB fad}) for Rayleigh fading ad hoc network with respect to the target SINR. Here $\mu=3.7$, $\alpha=10$ dB, and $\delta = -15$ dB. We also show analytical values calculated from the noise-free approximation (\ref{eq:SINR Q fad NoNoise}).}\label{fig:Outage rate fad}
\vspace{-0.1in}
\end{figure}

In Figs. \ref{fig:Avg rate fad} and \ref{fig:Outage rate fad}, we illustrate the average Shannon and outage rate over $\alpha$ and $\eta$, respectively, for Rayleigh fading ad hoc networks. For comparison, we also plot their analytical lower bounds given in Lemma \ref{lemma:Rate FD}, and the analytical values calculated based on the noise-free CDF approximation (\ref{eq:SINR Q fad NoNoise}). As expected, the bounds for the fading channel are much tighter than the non-fading counterpart given in Section \ref{sec:ResStoGeo NonFad}.

Similar to the non-fading case, here the CDF and rates are also independent of the density. The reason is the same as in Remark \ref{remark:1}. Finally, we note that given $\delta = 0$, the average of the inverted SIR $\mathbb{E} \left[ Q_{\text{fd}}^{-1} \right]$ goes to infinity since $\mathbb{E}_h \left[ \left| h \right|^{-2} \right]$ is unbounded, where the random variable $h \sim \mathcal{CN}(0,1)$.

\section{Performance of Partial Fading Ad-Hoc Networks}\label{sec:ResStoGeo ParFad}

In this section, we will investigate partial fading ad-hoc networks, under which only the interfering signals experience Rayleigh fading channels while the desired signal does not. The SINR of a typical user is expressed as
\begin{align}\label{eq:SINR Q parfad}
Q_{\text{p-fd}} \triangleq \frac{ d_1^{-\mu}}{\sum_{k\in \Phi \setminus \{1\}} \left| h_k \right|^2 d_k^{-\mu} + \delta},
\end{align}
where $\Phi$, $\left\{ h_k \right\}_{k=1}^{\infty}$, $\left\{ d_k \right\}_{k=1}^{\infty}$, and $\delta$ have been defined in Section \ref{sec:ad hoc netw}. In the literature, $Q_{\text{p-fd}}$ has also been studied in \cite{WeAnJi07,WeAnJi10}. However, as discussed in Section \ref{sec:ResStoGeo NonFad}, the distance $d_{1}$ is assumed to be a constant and only the set $\left\{ d_{k} \right\}_{k=2}^{\infty}$ follows a PPP \cite{WeAnJi07,WeAnJi10}. Here, we consider the case when $\left\{ d_{k} \right\}_{k=1}^{\infty}$ follows the PPP $\Phi$ with density $\lambda$ and $d_k \geq d_1$, $\forall k \geq 2$.

Similar to Section \ref{sec:ResStoGeo NonFad}, we will derive the Laplace transform of $Q_{\text{p-fd}}^{-1}$ and the distribution bounds for $Q_{\text{p-fd}}$ assuming that $\delta = 0$. The results are presented in the following lemmas. Note that no upper bound is available due to the difficulty of the derivation.

\begin{lemma}\label{lemma:Laplace ParFad}
The Laplace transform of $Q_{\text{p-fd}}^{-1}$ is given as
\begin{align}
& \mathcal{L}_{1/Q_{\text{p-fd}}} \left(s \right) = 2\pi \lambda \int_{0}^{\infty} x \exp \bigg\{ -s x^{\mu} \delta \notag \\
& \qquad - \pi \lambda \left[ x^2 + 2 \int_{x}^{\infty} \bigg( 1 - \frac{1}{1 + s x^{\mu}v^{-\mu}}  \bigg) dv \right]   \bigg\} dx
\end{align}
\end{lemma}
\begin{IEEEproof}
Please refer to Appendix \ref{appen:Lap ParFad}
\end{IEEEproof}

\begin{lemma}\label{lemma:ParFad CDF bounds}
Assume $\delta = 0$. The lower bound of the CDF of $Q_{\text{p-fd}}$ defined in (\ref{eq:SINR Q parfad}) is given as
\begin{align}\label{eq:ParFD LB}
F^{\text{UB}}_{Q_{\text{p-fd}}}(q) 
& = 1 - \frac{1}{  1 +  \frac{2 q^{2/\mu}}{\mu} \Gamma\left( \frac{2}{\mu}, \frac{1}{q}  \right)  }.
\end{align}
\end{lemma}
\begin{IEEEproof}
The proof follows from (\ref{eq:appen 2}) in Appendix \ref{appen:FD UB LB CDF}.
\end{IEEEproof}

It is difficult to derive a CDF upper bound of partial fading ad-hoc networks. In the following, we will present an interesting relationship between non-fading and partial fading networks which leads to such bound. Note that given a PPP $\Phi$ configuration, $\sum_{k\in \Phi \setminus \{1\}} \left| h_k \right|^2 d_k^{-\mu}$ is a summation of an infinite number of independent random variables. Without loss of generality, we can rank $\{ d_k \}_{k\in \Phi}$ as $d_1 \leq d_2 \leq ...$. Due to the Kolmogorov's strong law of large number \cite[Theorem 2.3.10]{SeSi93}, we have
\begin{align}
\frac{1}{N} \sum_{k=2}^N \left| h_k \right|^2 d_k^{-\mu} \xrightarrow{N \to \infty} \mathbb{E}_{ \frac{1}{N} \{ h_k \}} \bigg[ \sum_{k=2}^N \left| h_k \right|^2 d_k^{-\mu} \bigg] = \frac{1}{N} \sum_{k=2}^N d_k^{-\mu},
\end{align} 
since 
\begin{align}
\sum_{k=2}^{\infty} \frac{1}{k^2} \text{var}_{h_k} \Big[ \left| h_k \right|^2 d_k^{-\mu} \Big] < \sum_{k=2}^{\infty} d_k^{-\mu} < \infty
\end{align}
with high probability when $\mu >2$. Therefore, we can approximate
\begin{align}
\sum_{k\in \Phi \setminus \{1\}} \left| h_k \right|^2 d_k^{-\mu} \approx \sum_{k\in \Phi \setminus \{1\}} d_k^{-\mu},
\end{align}
and $Q_{\text{p-fd}} \approx Q_{\text{n-fd}}$. With this approximation, we can utilize all results, not only restrict to the upper bound, from Section \ref{sec:ResStoGeo NonFad} to study partial fading ad-hoc networks. 

\begin{figure}[t]
 		\centering 
 		\epsfxsize=0.48\linewidth
 		\captionsetup{width=0.48\textwidth} 
 		\includegraphics[width=8.5cm, height=7.5cm]{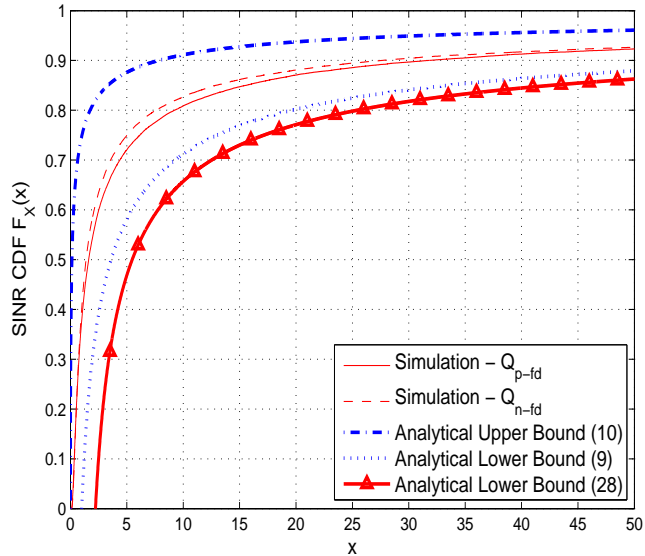}
 		\caption{Simulated SINR CDFs of $Q_{\text{p-fd}}$ and $Q_{\text{n-fd}}$ and their bounds with $\mu=3.7$ and $\delta=-20$ dB.}\label{fig:CDF par-fad}
\vspace{-0.1in}
\end{figure}

In Fig. \ref{fig:CDF par-fad}, we show the simulated SINR CDFs of non-fading and partial fading ad-hoc networks as well as their bounds. Here, the number of channel realizations is $20000$. It is observed that the CDFs of $Q_{\text{p-fd}}$ and $Q_{\text{n-fd}}$ are closely matched. Furthermore, even through simpler, the lower bound (\ref{eq:ParFD LB}) is much looser than (\ref{eq:LB all Q}. Together, those observations well justify the approximation $Q_{\text{p-fd}} \approx Q_{\text{n-fd}}$ and the use of the results in Section \ref{sec:ResStoGeo NonFad} to study $Q_{\text{p-fd}}$.

\section{Case Study: Massive MIMO and Small-Cell Densification in 5G}\label{sec:app and discuss}

Towards the development of the fifth generation (5G) communication systems, there emerge many promising technologies. It is therefore necessary to compare their performance under different scenarios, developing insights to deploy the right technology in the right scenarios. Note that among the technologies are M-MIMO and small-cell densification, both promising higher achievable throughput with either concentrating massive number of antennas in one location and serving the users with narrow beams, or distributing the large number of antennas close to the users' proximity.

The results obtained in Sections \ref{sec:ResStoGeo NonFad} and \ref{sec:ResStoGeo Fad} are useful to obtain performance insights for 5G systems, since many of them can be modeled or reduced to non-fading and Rayleigh fading ad-hoc networks. Furthermore, it is possible to compare the performance of such systems using the bounds given in Sections \ref{sec:ResStoGeo NonFad} and \ref{sec:ResStoGeo Fad}, due to their closed-form expressions.
 
Here, we give an example of M-MIMO networks. Consider the system model in Section \ref{sec:ResStoGeo NonFad} but with $M$ antennas at each BS. The base-band received signal for the UE in cell $b$ is given as
\begin{align}\label{eq:MMIMO example}
y_{b} = \mv{h}_{b,b}^H \mv{x}_{b} + \displaystyle \sum_{ b' \neq b }^{\infty} \mv{h}_{b',b}^H \mv{x}_{b'} + n_{b},
\end{align}
where $\mv{h}_{b',b}^H$ is the channel between BS $b'$ and the associated UE of BS $b$; and $\mv{x}_{b'}$ is the transmit signal from BS $b'$. Here, $\mv{h}_{b',b} = \sqrt{\xi_{b',b}} \widehat{\mv{h}}_{b',b}$; $\xi_{b',b}$ = $d_{b',b}^{-\mu}$ is a large scale fading including path loss with $\mu$ denoting the path loss exponent; $\widehat{h}_{b',b}$ follows a statistical distribution $\widehat{h}_{b',b} \sim \mathcal{CN}(0,1)$; and $n_{b}$ is the additive white Gaussian noise (AWGN) distributed as $\mathcal{CN}(0,\sigma_n^2)$.

Assume that each BS has a total power constraint of $MP_T$ and therefore $\mathbb{E}|\mv{x}_{b'}|^2 = M P_T$. Furthermore, assume that each BS employs the (optimal) conjugate beamforming, i.e., $\mv{x}_{b} = \mv{w}_{b} s_{b} = \frac{\sqrt{MP_T} \mv{h}_{b,b}}{\lVert \mv{h}_{b,b} \rVert} s_{b}$ where $\mv{w}_{b}$ and $s_{b}$ are the beamformer and desired signal of BS $b$, respectively. We can express the SINR of the UE associated to BS $b$ as 
\begin{align}\label{eq:SINR Asymp}
\text{SINR}_{b} 
&= \frac{ d_{b,b}^{-\mu}  \left\lVert \widehat{\mv{h}}_{b,b} \right\rVert^2/M } { \sum\limits_{ b' \neq b }^{\infty} \displaystyle d_{b',b}^{-\mu} \frac{ \widehat{\mv{h}}_{b',b}^H \widehat{\mv{h}}_{b',b'} \widehat{\mv{h}}_{b',b'}^H \widehat{\mv{h}}_{b',b} }{ M \left\lVert \widehat{\mv{h}}_{b',b'} \right\rVert^2 } + \frac{\sigma_n^2}{M^2P_T}} \notag \\
& \xrightarrow{M\to \infty} \frac{ M d_{b,b}^{-\mu}  } { \sum\limits_{ b' \neq b }^{\infty} \displaystyle d_{b',b}^{-\mu} |\tilde{h}_{b'}|^2 + \frac{\sigma_n^2}{MP_T}},
\end{align}
where $ \tilde{h}_{b'} \triangleq \frac{ \widehat{\mv{h}}_{b',b}^H \widehat{\mv{h}}_{b',b'} }{ \left\lVert \widehat{\mv{h}}_{b',b'} \right\rVert }$ are i.i.d. random variables each distributing as $\mathcal{CN}(0,1)$.

It can be observed that (\ref{eq:SINR Asymp}) has a similar form to (\ref{eq:SINR Q parfad}). We thus can apply the results in Sections \ref{sec:ResStoGeo NonFad} and \ref{sec:ResStoGeo ParFad} to investigate the performance of M-MIMO networks modeled as in (\ref{eq:MMIMO example}). Note that small-cell systems can be modeled as Rayleigh fading ad-hoc networks. Therefore, we can further compare the performance of M-MIMO and small cell networks due to the closed-form expressions of the performance bounds in Section \ref{sec:ResStoGeo NonFad} and \ref{sec:ResStoGeo Fad}. Please refer to our subsequent study \cite{NgSu15} for more details.

\section{Conclusions}\label{sec:conclusions}

In this paper, we have investigated the non-fading and Rayleigh fading ad hoc networks. In particular for the non-fading case, we have derived the Laplace transform of the inverted SINR, which is unfortunately complicated and of limited use. Therefore, we have also proposed upper and lower bounds for the SIR distribution and given the closed-form expressions. The results have then been used to obtain the bounds for both the average Shannon and outage rates. Similarly for Rayleigh fading and partial fading ad hoc networks, we have derived the bounds for the SIR distribution, average Shannon rate, and outage rate of the system. Since many communication networks can be modeled or reduced to these two ad hoc networks, our results are potentially useful for a wide range of problems. As an example, we have shown that our results can be applied to study and compare the performance of two 5G communication technologies, i.e., massive MIMO and small-cell densification.

\appendices

\section{Proof of Theorem \ref{theorem:GenLap}}\label{appen:GenLap}

We first derive the Laplace transform of the random variable $Q_{\text{n-fd}}^{-1}$ given $d_1 = x$ as follows. 
\begin{align}
& \mathcal{L}_{1/Q_{\text{n-fd}}} \left(s|d_{1}=x \right) = \mathbb{E}_{\Phi} \left[ e^{-sQ_{\text{n-fd}}} \right] \notag \\
& = \mathbb{E}_{\Phi} \left[ \exp \left( - s x^{\mu} \sum\limits_{ k=2 }^{\infty} \displaystyle d_{k}^{-\mu} - s x^{\mu} \delta \right)  \right] \notag \\
& = e^{-s x^{\mu} \delta} \mathbb{E}_{\Phi} \left[ \prod\limits_{ k=2 }^{\infty} e^{-s x^{\mu} d_{k}^{-\mu}} \right] \notag \\
& \overset{(a)}{=} e^{-s x^{\mu} \delta} \exp\left\{ - \int_{x}^{\infty} \left( 1- e^{-s x^{\mu} r^{-\mu}} \right) 2\pi \lambda r dr  \right\} \notag \\
& \overset{(b)}{=} e^{-s x^{\mu} \delta} \exp \bigg\{ - \pi \lambda x^2 (1-e^{-s}) \notag \\
& \qquad \qquad \qquad + \pi \lambda \mu s x^{\mu} \int_{x}^{\infty} r^{-\mu+1} e^{-s x^{\mu} r^{-\mu}} dr  \bigg\} \notag \\
& \overset{(c)}{=} e^{-s x^{\mu} \delta} \exp\bigg\{ - \pi \lambda x^2 (1-e^{-s}) \notag \\
& \qquad \qquad \qquad - \pi \lambda s x^{\mu} \int_{0}^{x^{-\mu}} v^{-2/\mu} e^{-s x^{\mu} v} dv  \bigg\} \notag \\
& \overset{(d)}{=} e^{-s x^{\mu} \delta} \exp\left\{ - \pi \lambda x^2 (1-e^{-s}) - \pi \lambda s^{2/\mu} x^{2} \gamma(1-2/\mu,s)  \right\},
\end{align}
where $(a)$, $(b)$, $(c)$, $(d)$ is due to the Campbell's theorem \cite{Ha12}, the integration by part, the change of variable $v = r^{-\mu}$, and \cite[(3.351.1)]{GrRy07}, respectively. Here $\gamma(\cdot,\cdot)$ is the lower incomplete gamma function, defined as $\gamma(a,b) = \int_{0}^{b} t^{a-1} e^{-t} dt$.

Now note that $d_{1}$ is the distance from an arbitrary origin to the nearest point of the PPP $\Phi$ with density $\lambda$. Therefore, the PDF of $d_{1}$ is given as \cite{AnBaGa11} \cite{Mo12}
\begin{align}\label{eq:PDFnearest}
f_{D_1}(x) = 2\pi \lambda x e^{-\pi \lambda x^2}. 
\end{align}

The Laplace transform of $Q_{\text{n-fd}}^{-1}$ is thus expressed as
\begin{align}
& \mathcal{L}_{1/Q_{\text{n-fd}}} \left(s \right) \notag \\
& = \int_{0}^{\infty} \mathcal{L}_{Q_{\text{n-fd}}} \left(s|d_{1}=x \right) f_{D_1}(x) dx \notag \\
& = 2\pi \lambda \int_{0}^{\infty} x \exp \bigg\{ -s x^{\mu} \delta \notag \\
& \qquad \qquad - \pi \lambda x^2 \left[ 2- e^{-s} + s^{2/\mu} \gamma\left( 1- \frac{2}{\mu},s \right)  \right]   \bigg\} dx \notag \\
& \overset{(e)}{=} \pi \lambda \int_{0}^{\infty}  \exp \bigg\{ -s y^{\mu/2} \delta \notag \\
& \qquad \qquad - \pi \lambda y \left[ 2- e^{-s} + s^{2/\mu} \gamma\left( 1- \frac{2}{\mu},s \right)  \right]   \bigg\} dy,
\end{align}
where $(e)$ is due to the change of variable $y = x^2$. 

This concludes the proof of Theorem \ref{theorem:GenLap}.

\section{Proof of Lemma \ref{lem:BouCDF g1}}\label{appen:BouCDF g1}

\subsection{Lower Bound}\label{appenSec:LowBou}

Considering $q\geq 1$ and the second smallest distance $d_2$ only, we have  
\begin{align}\label{eq:appen 1}
F_{Q_{\text{n-fd}}} (q|d_1=x) 
& \geq \text{Pr} \left\{ d_2 \leq x q^{1/\mu} |d_1=x \right\} \notag \\
& \overset{(a)}{=} \int_{x}^{x q^{1/\mu}} 2\pi \lambda t e^{-\pi\lambda (t^2-x^2)} dt \notag \\
& = 1 - e^{-\pi\lambda (q^{2/\mu} -1)x^2},
\end{align}
where $(a)$ is due to \cite[(28)]{Mo12}. The PDF of $d_1$ follows (\ref{eq:PDFnearest}), which thus leads to
\begin{align}\label{eq:appen FlbQ}
& F^{\text{LB}}_{Q_\text{n-fd}}(q) 
= \int_{0}^{\infty} F^{\text{LB}}_{Q_\text{n-fd}}(q|d_1=x) f_{D_1}(x) dx \notag \\
& = \int_{0}^{\infty} \left(  1 - e^{- \lambda \pi \left( q^{2/\mu}-1 \right) x^2} \right)2\pi \lambda x e^{-\pi \lambda x^2} dx 
= 1 - \frac{1}{q^{2/\mu}}.
\end{align}

\subsection{Upper Bound}\label{appenSec:UppBou}
Assuming that $q\geq 1$ and $\xi\geq 1$ is an arbitrary parameter, we now apply the dominant-interferer concept \cite{WeAnJi07}. The idea is to partition the set of interferers into dominating and non-dominating points, in which a point $k$ $(\geq 2)$ is called dominating if it alone can cause outage at the UE. The sets of dominating and non-dominating interferers, denoted as $\mathcal{S}_{\text{dom}}$ and $\mathcal{S}_{\text{n-dom}}$, respectively, are defined as follows
\begin{align}
\mathcal{S}_{\text{dom}} & = \left\{ k: k \geq 2, d_k \leq (\xi q)^{1/\mu} d_1  \right\}, \notag \\
\mathcal{S}_{\text{n-dom}} & = \left\{ k: k \geq 2, d_k > (\xi q)^{1/\mu} d_1  \right\}.
\end{align}
For ease of representation, we also define $\Sigma_{\text{n-dom}} = d_1^{\mu} \sum_{k \in \mathcal{S}_{\text{n-dom}}} d_k^{-\mu}$ and $\Sigma_{\text{dom}} = d_1^{\mu} \sum_{k \in \mathcal{S}_{\text{dom}}} \xi d_k^{-\mu}$. Note that
\begin{align}
& \widehat{Q}_{\text{n-fd}} \triangleq \frac{d_1^{-\mu}}{ \sum_{k \in \mathcal{S}_{\text{dom}}} \xi d_k^{-\mu} + \sum_{k \in \mathcal{S}_{\text{n-dom}}} d_k^{-\mu}} \notag \\
& \qquad \qquad \leq \frac{d_1^{-\mu}}{ \sum_{k \in \mathcal{S}_{\text{dom}}} d_k^{-\mu} + \sum_{k \in \mathcal{S}_{\text{n-dom}}} d_k^{-\mu}} = Q_{\text{n-fd}}.
\end{align}

We thus have 
\begin{align}
F_{Q_\text{n-fd}}(q) \leq \text{Pr} \left\{ \widehat{Q}_{\text{n-fd}} \leq q \right\}  \triangleq  F^{\text{UB}}_{Q_\text{n-fd}}(q). 
\end{align}

Given that $d_1 = x$, we have
\begin{align}\label{eq:appen 6}
& \text{Pr} \left\{ \widehat{Q}_{\text{n-fd}} \leq q |d_1=x \right\} \notag \\
& = \text{Pr} \bigg\{ \left\{ \Sigma_{\text{dom}} \geq \frac{1}{q} \right\} \cup \left\{ \Sigma_{\text{n-dom}} \geq \frac{1}{q} \right\}  \notag \\
& \quad \cup \left\{ \Sigma_{\text{dom}} < \frac{1}{q}, \Sigma_{\text{n-dom}} < \frac{1}{q}, \Sigma_{\text{dom}} + \Sigma_{\text{n-dom}} \geq \frac{1}{q} \right\} \Big|d_1=x \bigg\} \notag \\
& \overset{(a)}{=} \text{Pr} \left\{ \left\{ \Sigma_{\text{dom}} \geq \frac{1}{q} \right\} \cup \left\{ \Sigma_{\text{n-dom}} \geq \frac{1}{q} \right\} \Big|d_1=x \right\} \notag \\
& \overset{(b)}{=}   \text{Pr} \left\{ \Sigma_{\text{dom}} \geq \frac{1}{q} \Big|d_1=x \right\} +  \text{Pr} \left\{ \Sigma_{\text{n-dom}} \geq \frac{1}{q} \Big|d_1=x \right\}  \notag \\
& \qquad -  \text{Pr} \left\{ \Sigma_{\text{dom}} \geq \frac{1}{q} \Big|d_1=x \right\} \text{Pr} \left\{ \Sigma_{\text{n-dom}} \geq \frac{1}{q} \Big|d_1=x \right\} \notag \\
& = 
\text{Pr} \left\{ \Sigma_{\text{dom}} \geq \frac{1}{q} \Big|d_1=x \right\} + \left( 1 - \text{Pr} \left\{ \Sigma_{\text{dom}} \geq \frac{1}{q} \Big|d_1=x \right\} \right)  \notag \\
& \qquad \qquad \times \text{Pr} \left\{ \Sigma_{\text{n-dom}} \geq \frac{1}{q} \Big|d_1=x \right\}, 
\end{align}
where $(a)$ is due to the fact that the event $ \left\{ \Sigma_{\text{dom}} < 1/q |d_1=x \right\}$ is null; $(b)$ is because of the independent of $\mathcal{S}_{\text{dom}}$ and $\mathcal{S}_{\text{n-dom}}$. From (\ref{eq:appen 1}), we obtain  
\begin{align}\label{eq: appen 3}
\text{Pr} \left\{ \Sigma_{\text{dom}} \geq \frac{1}{q} \Big|d_1=x \right\} 
& = \text{Pr} \left\{ d_2 \leq x (\xi q)^{1/\mu} |d_1=x  \right\} \notag \\
& = 1 - e^{-\pi\lambda ((\xi q)^{2/\mu} -1)x^2}.
\end{align} 

Furthermore, we have
\begin{align}
\text{Pr} \left\{ \Sigma_{\text{n-dom}} \geq \frac{1}{q} \Big|d_1=x \right\} 
\overset{(c)}{\leq} q \mathbb{E}_{\Phi}\left[ \widehat{\Sigma}_{\text{n-dom}} |d_1=x \right], \label{eq: appen 4}
\end{align}
where $(c)$ is due to the Markov inequality $\text{Pr} \{X>a\}$ $\leq \frac{1}{a} \mathbb{E}_X[X]$. Now by applying Campbell's theorem, we further obtain
\begin{align}\label{eq: appen 2}
& \mathbb{E}_{\Phi} \left[ \Sigma_{\text{n-dom}} |d_1=x \right] 
= \mathbb{E}_{\Phi} \left[ x^{\mu} \sum_{k\in \mathcal{S}_{\text{n-dom}}} d_k^{-\mu} \right] \notag \\
& \qquad =   x^{\mu} \int_{(\xi q)^{1/\mu} x}^{\infty} v^{-\mu} 2\pi \lambda v dv 
= \frac{2\pi \lambda x^2 (\xi q)^{2/\mu-1}}{\mu-2}.
\end{align}

Substituting (\ref{eq: appen 2}) and (\ref{eq: appen 3}) to (\ref{eq: appen 4}), the upper bound given that $d_1=x$ is expressed as follows 
\begin{align}\label{eq: appen 5}
& F^{\text{UB}}_{Q_\text{n-fd}}(q|d_1=x) 
= 1 - e^{- \lambda \pi \left( (\xi q)^{2/\mu}-1 \right) x^2} \notag \\
& \qquad \qquad + e^{ - \lambda \pi \left( (\xi q)^{2/\mu}-1 \right) x^2} \frac{2\pi \lambda x^2 \xi^{2/\mu-1} q^{2/\mu}}{\mu-2},
\end{align}
and consequently we have 
\begin{align}
& F^{\text{UB}}_{Q_\text{n-fd}}(q) 
= \int_{0}^{\infty} F^{\text{UB}}_{Q_\text{n-fd}}(q|d_1=x) f_{D_1}(x) dx \notag \\
& = 2\pi \lambda \int_{0}^{\infty} \left(  1 - e^{- \lambda \pi \left( (\xi q)^{2/\mu}-1 \right) x^2} \right) x e^{-\pi \lambda x^2} dx \notag \\
& \qquad \qquad + \frac{ (2\pi \lambda)^2 \xi^{2/\mu-1} q^{2/\mu}}{\mu-2} \int_{0}^{\infty} x^3 e^{-\pi \lambda (\xi q)^{2/\mu}x^2} dx  \notag \\
& \overset{(e)}{=} 1 - \frac{1}{(\xi q)^{2/\mu}} + \frac{2 \xi^{2/\mu-1} q^{2/\mu}}{\mu-2} \frac{1}{(\xi q)^{4/\mu}} \notag \\
& = 1 - \frac{(\mu-2) \xi - 2}{(\mu-2)\xi^{1+2/\mu}} \frac{1}{q^{2/\mu}},
\end{align}
where $(e)$ is due to (\ref{eq:appen FlbQ}) and \cite[(3.321.6)]{GrRy07}. The final step is to find the maximum of $\frac{(\mu-2) \xi - 2}{(\mu-2)\xi^{1+2/\mu}}$ given that $\xi\geq 1$. The optimal $\xi^*$ is $\frac{\mu+2}{\mu-2}$. The result is (\ref{eq:UB Q g1}).

This completes the proof of Lemma \ref{lem:BouCDF g1}.

\section{Proof of Lemma \ref{lem:BouCDF l1}}\label{appen:BouCDF l1}
Using the Markov inequality, we first have
\begin{align}
F_{Q_{\text{n-fd}}}(q| d_1=x) 
& = \text{Pr} \left\{ Q_{\text{n-fd}}^{-1} \geq q^{-1} | d_1=x\right\} \notag \\
& \leq q \mathbb{E}_{\Phi} \left[ Q_{\text{n-fd}}^{-1} | d_1=x \right]  
= \mathbb{E}_{\Phi} \left[ x^{\mu} \sum_{k\geq 2} d_k^{-\mu} \right] \notag \\
& =  q x^{\mu} \int_{x}^{\infty} v^{-\mu} 2\pi \lambda v dv = \frac{2\pi \lambda q }{\mu-2} x^2.
\end{align}

The CDF is thus bounded above as follows
\begin{align}
F_{Q_{\text{n-fd}}}(q) \leq \int_{0}^{\infty} \frac{2\pi \lambda q }{\mu-2} x^2 2\pi \lambda x e^{-\pi\lambda x^2} dx
= \frac{2}{\mu-2} q.
\end{align}

Before proceeding, we note that $\frac{2}{\mu-2} q$ does not give a strict bound for $q \geq 1$ or $q< 1$ but $q\to 1$, since $1 - \frac{\mu (\mu-2)^{2/\mu}}{(\mu+2)^{2/\mu+1}} < \frac{2}{\mu-2}$. Now it is observed that, as $q\to 0$,
\begin{align}
\frac{2}{\mu-2} q < 1 - \frac{1}{1 + \beta_{\text{n-fd}} q^{2/\mu}}.
\end{align}

Furthermore based on Lemma \ref{lem:BouCDF g1}, it is straightforward to show that $1 - \frac{1}{1 + \beta_{\text{n-fd}} q^{2/\mu}}$ is an upper bound when $q\to 1$. 

This concludes the proof of Lemma \ref{lem:BouCDF l1}.

\section{Proof of Lemma \ref{lemma:Ravg UB LB}}\label{appen:Ravg UB LB}

We have
\begin{align}\label{eq:appen 3}
R_{\text{n-fd}}(\alpha) 
& = \int_{0}^{\infty} \text{Pr} \left\{ Q \geq \frac{2^t - 1}{\alpha}  \right\} dt \notag \\
& = \int_{0}^{\log_2(1+\alpha)} \left( 1 - \text{Pr} \left\{ Q \leq \frac{2^t - 1}{\alpha}  \right\} \right) dt \notag \\
& \quad + \int_{\log_2(1+\alpha)}^{\infty} \left( 1 - \text{Pr} \left\{ Q \leq \frac{2^t - 1}{\alpha}  \right\} \right) dt.
\end{align}

Substituting (\ref{eq:LB all Q}) to (\ref{eq:appen 3}), we have
\begin{align}
R_{\text{n-fd}} 
& \leq R^{\text{UB}}_{\text{n-fd}} 
= \int_{0}^{\log_2(1+\alpha)} dt + \int_{\log_2(1+\alpha)}^{\infty} \frac{\alpha^{2/\mu} dt}{\left( 2^t - 1\right)^{2/\mu} } \notag \\
& \overset{(a)}{=} \log_2(1+\alpha) + \frac{\alpha^{2/\mu}}{\left( 1 - \frac{2}{\mu} \right) \log 2} \int_{\log_2(1+\alpha)}^{\infty} \frac{dv}{ v^{\frac{1}{1-2/\mu}} + 1 }  \notag \\
& = \log_2(1+\alpha) + \frac{\alpha^{2/\mu} \mu}{(\mu-2)\log 2} \rho \left( \alpha^{1-2/\mu} , \frac{\mu}{\mu-2} \right),
\end{align}
where $(a)$ is due to the change of variable $v = \left( 2^t - 1 \right)^{1-2/\mu}$. The lower bounds can be obtained with similar steps by substituting (\ref{eq:UB all Q}) into (\ref{eq:appen 3}), and thus are omitted for brevity.

This completes the proof of Lemma \ref{lemma:Ravg UB LB}.

\section{Proof of Lemma \ref{lemma:FD UB LB CDF}}\label{appen:FD UB LB CDF}

\subsection{Lower Bound}

Given $h_1$ and $d_1$, we first define
\begin{align}
\mathcal{S}_{\text{fd,dom}} = \left\{ k \geq 2:  |h_k|^2 d_k^{-\mu} \geq q^{-1} |h_1|^2 d_1^{-\mu}  \right\}.
\end{align}

The SINR CDF can be bounded below as
\begin{align}
F_{Q_{\text{fd}}}(q|d_1,h_1) \geq F^{\text{LB}}_{Q_{\text{fd}}}(q|d_1,h_1),
\end{align}
where
\begin{align}
& F^{\text{LB}}_{Q_{\text{fd}}}(q|d_1,h_1) 
\triangleq  \text{Pr} \left\{ |h_1|^{-2} d_1^{\mu} \sum_{k \in \mathcal{S}_{\text{fd,dom}}} |h_k|^2 d_k^{-\mu} \geq q^{-1}  \right\} \notag \\
& \overset{(a)}{=} 1 - \text{Pr} \left\{  \mathcal{S}_{\text{fd,dom}} = \emptyset  \right\} \notag \\
& \overset{(b)}{=} 1 - \exp\left( - \int  \text{Pr} \left\{  |h|^2 |x|^{-\mu} \geq q^{-1} |h_1|^2 d_1^{-\mu}  \right\} dx \right),
\end{align}
where $(a)$ is due to the fact that the event $\left\{ \mathcal{S}_{\text{fd,dom}} \neq \emptyset  \right\}$ is the same as $\left\{  \sum_{k \in \mathcal{S}_{\text{fd,dom}}} |h_k|^2 d_k^{-\mu} \geq q^{-1} |h_1|^2 d_1^{-\mu} \right\}$; and $(b)$ is due to the void probability of a Poisson process. Using the polar coordinates and the fact that $|h|^2$ is exponentially distributed, we have
\begin{align}
& \int  \text{Pr} \left\{  |h|^2 |x|^{-\mu} \geq q^{-1} |h_1|^2 d_1^{-\mu}  \right\} dx \notag \\ 
& = \int e^{q^{-1} |h_1|^2 d_1^{-\mu} |x|^{\mu}} dx 
= \int_{d_1}^{\infty} e^{q^{-1} |h_1|^2 d_1^{-\mu} r^{\mu}} 2 \pi \lambda r dr \notag \\
& \overset{(c)}{=} \frac{2\pi \lambda}{\mu} \frac{d_1^2 q^{2/\mu}}{|h_1|^{2/\mu}} \Gamma\left( \frac{2}{\mu}, \frac{|h_1|^2}{q}  \right),
\end{align}
where $(c)$ is due to \cite[(3.381.9)]{GrRy07}. The CDF lower bound thus can be obtained as
\begin{align}\label{eq:appen 2}
F^{\text{LB}}_{Q_{\text{fd}}}(q)
& = 1 - \mathbb{E}_{h_1,d_1} \left[  \exp \left( - \frac{2\pi \lambda}{\mu} \frac{d_1^2 q^{2/\mu}}{|h_1|^{2/\mu}} \Gamma\left( \frac{2}{\mu}, \frac{|h_1|^2}{q}  \right)  \right)   \right] \notag \\
& = 1 - \mathbb{E}_{h_1} \bigg[  \int_{0}^{\infty} \exp \left( - \frac{2 q^{2/\mu}}{\mu |h_1|^{2/\mu}} \Gamma\left( \frac{2}{\mu}, \frac{|h_1|^2}{q}  \right) \pi \lambda t^2 \right)  \notag \\
& \qquad \qquad \qquad \qquad \times 2\pi \lambda t e^{-\pi\lambda t^2} dt  \bigg] \notag \\
& = 1 - \mathbb{E}_{h_1} \left[  \frac{1}{  1 +  \frac{2 q^{2/\mu}}{\mu |h_1|^{2/\mu}} \Gamma\left( \frac{2}{\mu}, \frac{|h_1|^2}{q}  \right)  }  \right].
\end{align}

\subsection{Upper Bound}

We note that 
\begin{align}
\int_{q^{-2/\mu}}^{\infty} \frac{du}{1+u^{\mu/2}} 
& < \int_{0}^{\infty} \frac{du}{1+u^{\mu/2}} 
\overset{(d)}{=} \frac{2\pi}{\mu \sin\left( \frac{2\pi}{\mu} \right)} \notag \\
& = \Gamma\left( 1 + \frac{2}{\mu} \right) \Gamma\left( 1 - \frac{2}{\mu} \right),
\end{align}
where $(d)$ is from \cite[(3.241.2)]{GrRy07}. As a consequence, we have
\begin{align}
& F_{Q_{\text{fd}}}(q) = 1 - \frac{1}{1 + q^{2/\mu} \int_{q^{-2/\mu}}^{\infty} \frac{du}{1+u^{\mu/2}} } \notag \\
& \qquad \qquad \qquad \qquad \leq 1 - \frac{1}{1 + \beta_{\text{fd}} q^{2/\mu}}  \triangleq F^{\text{UB}}_{Q_{\text{fd}}}(q).
\end{align}

This completes the proof of Lemma \ref{lemma:FD UB LB CDF}.

\section{Proof of Lemma \ref{lemma:Laplace ParFad}}\label{appen:Lap ParFad}

We first derive the Laplace transform of the random variable $Q_{\text{p-fd}}^{-1}$ given $d_1 = x$ as follows. 
\begin{align}
& \mathcal{L}_{1/Q_{\text{p-fd}}} \left(s|d_{1}=x \right) = \mathbb{E}_{\Phi, \{h_k\}} \left[ e^{-sQ_{\text{p-fd}}} \right] \notag \\
& = \mathbb{E}_{\Phi, \{h_k\}} \left[ \exp \left( - s x^{\mu} \sum\limits_{ k=2 }^{\infty} \left| h_k \right|^2 \displaystyle d_{k}^{-\mu} - s x^{\mu} \delta \right)  \right] \notag \\
& = e^{-s x^{\mu} \delta} \mathbb{E}_{\Phi, \{h_k\}} \left[ \prod\limits_{ k=2 }^{\infty} e^{-s x^{\mu} \left| h_k \right|^2 d_{k}^{-\mu}} \right] \notag \\
& \overset{(a)}{=} \exp \bigg( -s \delta x^{\mu} - 2\pi \lambda \int_{x}^{\infty} \bigg( 1 - \frac{1}{1 + s x^{\mu}v^{-\mu}}  \bigg) dv \bigg),
\end{align}
where $(a)$ is due to \cite[(39)]{AnBaGa11}.

Again, since the PDF of $d_{1}$ is given as in (\ref{eq:PDFnearest}), the Laplace transform of $Q_{\text{p-fd}}^{-1}$ is thus expressed as
\begin{align}
& \mathcal{L}_{1/Q_{\text{p-fd}}} \left(s \right) \notag \\
& = \int_{0}^{\infty} \mathcal{L}_{Q_{\text{p-fd}}} \left(s|d_{1}=x \right) f_{D_1}(x) dx \notag \\
& = 2\pi \lambda \int_{0}^{\infty} x \exp \bigg\{ -s x^{\mu} \delta \notag \\
& \qquad \qquad - \pi \lambda \left[ x^2 + 2 \int_{x}^{\infty} \bigg( 1 - \frac{1}{1 + s x^{\mu}v^{-\mu}}  \bigg) dv \right]   \bigg\} dx.
\end{align}

This concludes the proof of Lemma \ref{lemma:Laplace ParFad}.

\balance



\end{document}